\documentclass[twocolumn]{aastex631}
\usepackage{ulem}
\hypersetup{linkcolor=red,citecolor=blue,filecolor=cyan,urlcolor=magenta}
\usepackage{amsmath}
\usepackage{subfigure}
\usepackage{float}
\usepackage{tensor}     
\usepackage{graphicx}   
\usepackage{bm}         
\usepackage{xcolor}     
\usepackage{color}      
\usepackage{longtable}  
\usepackage[section]{placeins} 
\usepackage[utf8]{inputenc} 
\newcommand{\nc}{\newcommand*} 

\nc{\al}{\alpha}
\nc{\s}{\sigma}
\nc{\dt}{\delta}
\nc{\Dt}{\Delta}
\nc{\Ld}{\Lambda}
\nc{\p}{\partial}
\nc{\om}{\omega}
\nc{\Om}{\Omega}
\nc{\rd}{\mathrm{d}}
\nc{\Od}[1]{\mathcal{O}(#1)} 
\nc{\kp}{\kappa}

\def\({\left(}
\def\){\right)}
\def\[{\left[}
\def\]{\right]}
\def\e{\begin{equation}}
\def\q{\end{equation}}
\def\m{\begin{eqnarray}}
\def\n{\end{eqnarray}}
\nc{\Eq}[1]{Eq.~\eqref{#1}}     
\nc{\Fig}[1]{Fig.~\ref{#1}}     
\nc{\Table}[1]{Table~\ref{#1}}  
\nc{\Sec}[1]{Sec.~\ref{#1}}     
\nc{\Msun}{M_\odot}             
\nc{\fpbh}{f_{\mathrm{pbh}}}    
\nc{\fpbhn}{f_{\mathrm{pbh0}}}    
\nc{\mR}{\mathcal{R}} 
\nc{\seq}{\sigma_{\mathrm{eq}}}
\nc{\ogw}{\Omega_{\mathrm{GW}}}
\nc{\gpcyr}{\mathrm{Gpc}^{-3}\,\mathrm{yr}^{-1}}
\nc{\lvc}{LIGO/Virgo} 
\nc{\SNR}{\mathrm{SNR}} 
\nc{\mmin}{{m_{\mathrm{min}}}}
\nc{\mmax}{{m_{\mathrm{max}}}}
\nc{\Mmin}{{M_{\mathrm{min}}}}
\nc{\fmin}{{f_{\mathrm{min}}}}
\nc{\VT}{\mathrm{VT}}
\nc{\rhoGW}{\rho_{\mathrm{GW}}}
\nc{\vth}{\vec{\theta}}
\nc{\vd}{\vec{d}}
\nc{\vla}{\vec{\lambda}}
\nc{\Nobs}{N_{\mathrm{obs}}}
\nc{\av}[1]{\langle #1 \rangle} 
\nc{\km}{\mathrm{km}}
\nc{\Mpc}{\mathrm{Mpc}}
\nc{\Tobs}{T_{\mathrm{obs}}}
\nc{\Ntemp}{N_{\mathrm{temp}}}
\nc{\uni}{\mathrm{U}}
\nc{\logu}{\operatorname{\mathrm{log-U}}}
\nc{\addref}{[\textcolor{red}{add ref}] } 
\nc{\eg}{\textit{e.g.~}}
\nc{\app}{\approx}
\nc{\hf}{\frac{1}{2}}
\nc{\discuss}{\textcolor{red}{Add discussion here!}}
\nc{\red}[1]{\textcolor{red}{#1}}
\nc{\Gm}{\Gamma}
\nc{\mH}{\mathcal{H}}
\nc{\cs}{c_s^2}
\nc{\Sij}[1]{S_{ij}^{(#1)}}
\nc{\vi}[1]{v_i^{(#1)}}
\nc{\no}{\nonumber}
\def\<{\left\langle}
\def\>{\right\rangle}

\nc{\bk}{\bm{k}}
\nc{\bq}{\bm{q}}
\nc{\bp}{\bm{p}}
\nc{\bl}{\bm{l}}
\nc{\bx}{\bm{x}}
\nc{\be}{\bm{\epsilon}}
\nc{\mS}{\mathcal{S}}
\nc{\te}{\tilde{\eta}}
\nc{\tp}{\tilde{p}}
\nc{\tk}{\tilde{k}}
\nc{\tx}{\tilde{x}}
\nc{\tF}{\tilde{F}}
\nc{\tA}{\tilde{A}}
\nc{\mkpq}{|\bk-\bp-\bq|}
\nc{\mpq}{|\bp-\bq|}
\nc{\mkp}{|\bk-\bp|}
\nc{\mSi}[1]{\mS^{(#1)}({\bk, \eta})}
\nc{\vk}{\vec{k}}
\nc{\kstar}{k_*}
\nc{\fstar}{f_*}
\nc{\xstar}{x_*}
\nc{\mpbh}{m_{\rm{pbh}}}
\nc{\bn}[1]{\dt\bm{t}_{\text{#1}}}
\nc{\bC}[1]{\bm{C}_{\text{#1}}}
\nc{\NTOA}{N_{\text{TOA}}}
\nc{\Nmode}{{N_{\text{mode}}}}
\nc{\ARN}{A_{\rm{RN}}}
\nc{\gRN}{\gamma_{\rm{RN}}}
\nc{\bS}{\mathbf{\Sigma}}
\nc{\br}{\mathbf{r}}
\nc{\bN}{\mathbf{R}}
\nc{\bbt}{\mathbf{t}}
\nc{\bth}{\mathbf{\theta}}
\nc{\bep}{\mathbf{\epsilon}}
\nc{\RN}{\mathrm{RN}}
\nc{\BN}{\mathrm{BN}}
\nc{\GN}{\mathrm{GN}}
\nc{\mcN}{\mathcal{N}}
\nc{\GWB}{\mathrm{GW}}
\nc{\yr}{\mathrm{yr}}
\nc{\Am}{\mathcal{A}}
\nc{\Dm}{\mathcal{D}}
\nc{\Hm}{\mathcal{H}}
\nc{\mrm}{\mathrm}
\nc{\BF}{\mathcal{BF}}
\nc{\bt}{\mathbf{t}}
\nc{\bd}{\mathbf{d}}
\nc{\ba}{\mathbf{a}}
\nc{\bnu}{\mathbf{\nu}}

\received{}
\revised{}
\accepted{}
\published{}
\begin{document}
	
\title{Constraining the Graviton Mass with the NANOGrav 15-Year Data Set}

\author{Yu-Mei Wu}
\email{ymwu@ucas.ac.cn} 
\affiliation{School of Fundamental Physics and Mathematical Sciences, Hangzhou Institute for Advanced Study, UCAS, Hangzhou 310024, China}
\affiliation{School of Physical Sciences, University of Chinese Academy of Sciences, No. 19A Yuquan Road, Beijing 100049, China}

\author{Zu-Cheng Chen}
\email{ zucheng.chen@bnu.edu.cn}
\affiliation{Department of Astronomy, Beijing Normal University, Beijing 100875, China}
\affiliation{Advanced Institute of Natural Sciences, Beijing Normal University, Zhuhai 519087, China}
\affiliation{Department of Physics and Synergistic Innovation Center for Quantum Effects and Applications, Hunan Normal University, Changsha, Hunan 410081, China}

\author{Yan-Chen Bi}
\email{biyanchen@itp.ac.cn}
\affiliation{School of Physical Sciences, 
    University of Chinese Academy of Sciences, 
    No. 19A Yuquan Road, Beijing 100049, China}
\affiliation{CAS Key Laboratory of Theoretical Physics, 
    Institute of Theoretical Physics, Chinese Academy of Sciences, Beijing 100190, China}
\author{Qing-Guo Huang}
\email{huangqg@itp.ac.cn}
\affiliation{School of Fundamental Physics and Mathematical Sciences, Hangzhou Institute for Advanced Study, UCAS, Hangzhou 310024, China}
\affiliation{School of Physical Sciences, 
    University of Chinese Academy of Sciences, 
    No. 19A Yuquan Road, Beijing 100049, China}
\affiliation{CAS Key Laboratory of Theoretical Physics, 
    Institute of Theoretical Physics, Chinese Academy of Sciences,
    Beijing 100190, China}

\begin{abstract}
The recently detected stochastic signal by several pulsar timing array collaborations, offers an opportunity to scrutinize the fundamental properties of gravity, including the potential mass of the graviton. In this study, we analyze the NANOGrav 15-year data set to search for a stochastic gravitational wave background with modified Hellings-Downs correlations predicted by massive gravity. While the Bayesian analysis comparing the massive gravity to massless gravity within the effective searchable mass range of $m_g\in [3\times 10^{-25}, 8 \times 10^{-24}]\,\rm{eV}/c^2$ does not yield an explicit upper bound as all the Bayes factors are smaller than $3$, the combined consideration of the minimum frequency inherent in a massive gravity and the observed spectrum leads to an upper limit of $m_g<8.2\times 10^{-24}\,\rm{eV}/c^2$.
\end{abstract}


\section{Introduction}
After the successful observation of gravitational waves from compact binary coalescences by ground-based interferometers~\citep{LIGOScientific:2016aoc}, pulsar timing arrays, considered the most promising instruments for the first detection of a stochastic gravitational-wave background (SGWB), are expected to achieve the next major breakthrough in the field of gravitational wave detection in the coming years~\citep{Taylor:2015msb,Burke-Spolaor:2018bvk}.
Pulsar timing arrays consist of highly stable millisecond pulsars, whose emitted pulses are monitored in terms of their arrival times to  discern the imprints of gravitational waves~\citep{1978SvA....22...36S,Detweiler:1979wn,1990ApJ...361..300F}. Specifically, an SGWB manifests as distinctive inter-pulsar correlated timing residuals, which represent the differences between actual and expected pulse arrival times. These correlations are referred to as Hellings-Downs correlations~\citep{Hellings:1983fr}.
After decades of observations involving dozens of pulsars, several major international pulsar timing array collaborations, including the North American Nanohertz Observatory for Gravitational Waves (NANOGrav, \cite{McLaughlin:2013ira}), the European PTA (EPTA, \cite{Kramer:2013kea}) along with the Indian PTA (InPTA,~\cite{2018JApA...39...51J}), Chinese PTA (CPTA, \cite{2016ASPC..502...19L}), and the Parkes PTA (PPTA, \cite{Manchester:2012za}), have recently achieved groundbreaking progress. They have found evidence for a stochastic signal with the  Hellings-Downs correlations in their latest data sets~\citep{NANOGrav:2023hde,NANOGrav:2023gor,Zic:2023gta,Reardon:2023gzh,Antoniadis:2023lym,Antoniadis:2023ott,Xu:2023wog}, pointing to the gravitational-wave origin of the signal.

Gravitational waves play a pivotal role as a fundamental tool for testing the theory of gravity, providing profound insights into important inquiries, such as the question of whether gravitons possess mass. This inquiry spans from the discovery of orbital period variations in binary pulsars~\citep{Hulse:1974eb} that served as indirect evidence of gravitational waves~\citep{Weisberg:1981mt}, to the direct confirmation of gravitational waves from binary black hole coalescences~\citep{LIGOScientific:2016aoc}. For instance, the binary pulsar PSRB1913+16 has established an upper bound as $7.6\times 10^{-20}\,\rm{eV}/c^2$ \citep{Finn:2001qi}, the first observed gravitational-wave event GW150914 has put the bound as $m_g \lesssim 1.2 \times 10^{-22}\,\rm{eV}/c^2$ \citep{LIGOScientific:2016lio}, and the latest gravitational-wave transient catalog, GWTC-3, has further improved the bound as $m_g \lesssim 1.27\times 10^{-23}\,\rm{eV}/c^2$ \citep{LIGOScientific:2021sio}.

A massive gravity theory would also yield predictions of an SGWB different from those of General Relativity~(GR) , enabling the testing of massive gravity theories using pulsar timing arrays~\citep{Lee:2010cg}. If gravitons possess mass, it would lead to a modification of the dispersion relation for gravitational waves, resulting in two notable consequences for the SGWB. Firstly, there would exist a lower limit on the frequency of gravitational waves. Secondly, this alteration in the dispersion relation would influence the propagation equation, affecting pulse arrival times and consequently generating correlations among different pulsars that deviate from the Hellings-Downs curve~\citep{Liang:2021bct}. 

Previous studies have constrained the graviton mass by fitting data to the correlation function~\citep{Bernardo:2023mxc, Wang:2023div}. However, in this work, we will take a different approach by directly searching for the SGWB from massive gravity using the NANOGrav 15-year data set.

\section{\label{sec:orf}SGWB from massive gravity}
By combining the de Broglie relations and the mass-energy equation, the component gravitational-wave signal of a massive SGWB can be described by a four-wave vector $k^{\mu}=(\omega/c, k)$ that satisfies, 
\e
\frac{\omega}{c}=\sqrt{\frac{m_g^2 c^2}{\hbar^2}+\lvert\bf{k}\rvert^2},
\label{w_k}
\q
where $\omega$ is the circular frequency,  $c$ is the speed of light, and $\hbar$ is the reduced Planck constant. The above relationship clearly demonstrates that there exists a minimal frequency for the gravitational-wave signal in a massive gravity, i.e.,
\e
f_{\text{min}}=\frac{m_g c^2}{2\pi \hbar}.
\q

For an SGWB formed by the superposition of gravitational-wave components of different frequencies, the induced timing residuals can be described by the cross-power spectral density,
\m
S_{ab}(f) = \frac{H_0^2}{16\pi^4 f^5}\Gm_{ab}(f)\Omega_{\text{gw}}(f), 
\label{psd}
\n
where $H_0$ is the Hubble constant,  $\Gm_{ab}$ is the overlap reduction function (ORF) that measures the spatial correlation between the pulsar pairs $a$ and $b$, and $\Omega_{\text{gw}}(f)$ is the dimensionless gravitational-wave energy density parameter. While a large population of inspiraling supermassive black hole binaries are typically considered the most anticipated source for the stochastic signal, there are numerous other astrophysical and cosmological explanations that can also account for it~\citep{NANOGrav:2023hvm, Wu:2023hsa, Liu:2023ymk,Bi:2023tib,Liu:2023pau,Jin:2023wri,Ellis:2023oxs,Yi:2023npi,Chen:2023zkb}. Given some remaining uncertainty in the spectral shape, we adopt the commonly-assumed power-law form for $\Omega_{\text{gw}}(f)$,
\m
\Omega_{\text{gw}}(f)=\frac{2\pi^2 A_{\text{gw}}^2}{3H_0^2}\(\frac{f}{f_{\text{yr}}}\)^{5-\gamma_{\text{gw}}}f_{\text{yr}}^{2},
\n
where $A_{\text{gw}}$ is the amplitude of the SGWB at the reference frequency $f_{\text{yr}}=1/\text{year}$ and $\gamma_{\text{gw}}$ is the spectral index.

When considering the dispersion relation described by  \Eq{w_k} in the context of massive gravity, the previously known massless Hellings-Downs correlation undergoes modification, resulting in the following expression:
\e
\begin{split}
\Gm_{\rm{MG}}^{ab} = & \frac{1}{16 \rm{\eta}^5}
 \bigg [2\rm{\eta}(3+(6-5\rm \eta^2)\delta)\\
 &-6 \left(1+\delta +\rm{\eta}^2(1-3\delta )\right)\ln \left(\frac{1+\rm{\eta}}{1-\rm{\eta}}\right)\\ 
&\left.-\frac{3 \left(1+2 \rm{\eta}^2(1-2\delta)-\rm{\eta}^4 (1-\delta^2 ) \right) \ln I}{ \sqrt{(1-\delta ) \left(2-\rm{\eta}^2 (1+\delta) \right)}}\right],
\end{split}
\label{orf2}
\q
where
\e
\begin{split}
I=&\frac{1}{\left(\rm{\eta}^2-1\right)^2}\bigg[1+2 \rm{\eta}^2(1-2\delta)-\rm{\eta}^4 (1-2\delta^2)\\
&-2 \rm{\eta}(1-2 \rm{\eta}^2 \delta )  \sqrt{(1-\delta ) \left(2-\rm{\eta}^2 (1+\delta) \right)}\bigg].
\end{split}
\q
In these equations, $\delta\equiv \cos{\xi}$, where $\xi$ represents the separation angle of the two pulsars, and $\eta\equiv c \lvert\bf{k}\rvert/ \omega$. When $\eta=1$, the gravitons become massless, and the above expressions reduce to the familiar Hellings-Downs function. However, it's worth noting that this reduction may not be immediately apparent from \Eq{orf2}, as it appears to diverge when $\eta=1$. In such cases, one can refer to Eq.~(6) in \cite{Wu:2023pbt}, which provides an alternative analytical form of \Eq{orf2} suitable for approximations when $\eta \approx 1$.

\section{\label{sec:data}The data set and methodology}
\begin{table*}[tbp]
    \centering
    \caption{Parameters and their prior distributions used in the analyses.}
    \label{prior}
    \begin{tabular}{c c c c}
        \hline
        \textbf{Parameter} & \textbf{Description} & \textbf{Prior} & \textbf{Comments} \\
        \hline
        \multicolumn{4}{c}{White Noise}\,\\	        
        $E_{k}$ & EFAC per backend/receiver system & $\uni[0.01, 10]$ & single pulsar analysis only \\
        $Q_{k}$[s] & EQUAD per backend/receiver system & $\logu[-8.5, -5]$ & single pulsar analysis only \\
        $J_{k}$[s] & ECORR per backend/receiver system & $\logu[-8.5, -5]$ & single pulsar analysis only \\
        \hline
        \multicolumn{4}{c}{Red Noise} \\
        $A_{\rm{RN}}$ & red-noise power-law amplitude &$\logu[-20, -11]$ & one parameter per pulsar\, \\
        $\gamma_{\rm{RN}}$ & red-noise power-law index  &$\uni[0,7]$ & one parameter per pulsar\, \\
        \hline
        \multicolumn{4}{c}{Common-spectrum Process}\,\\
        $A_{\mrm{gw}}$ & SGWB power-law amplitude   &$\logu[-18, -11]$ & one parameter per PTA\, \\
        $\gamma_{\mrm{gw}}$ & SGWB power-law spectral index  &$\uni[0,7]$ & one parameter per PTA\, \\
        $m_g~[\rm{eV}/c^2]$ & graviton mass & delta function in $[10^{-24.5}, 10^{-23.1}]$ & $m_g \in \{10^{-24.5},10^{-24.4},\dots,10^{-23.1}\}$ \, \\
        \hline
    \end{tabular}
\end{table*}


The NANOGrav 15-year data set comprises observations from 68 pulsars, of which 67 have an observational timespan exceeding 3 years and have been used in the search for the SGWB predicted by GR~\citep{NANOGrav:2023gor}. 

In this study, we will also utilize these 67 pulsars to search for the massive SGWB in their timing data.
The expected arrival times of pulses from one pulsar are described by a timing model that encompasses various astrometric and timing parameters, including the pulsar's position, proper motion and spin period. After subtracting the timing model from the actual timing data, one obtains the timing residuals. In practice, apart from the SGWB signal, several effects will also contribute to timing residuals, including the inaccuracies of the timing model, other stochastic processes which can be further categorized as red noise and white noise. 
Following \cite{NANOGrav:2023gor}, the timing residuals $\dt \bbt $ for each pulsar can be decomposed into
\m
\dt \bbt=\bbt_{\rm{TM}} + \dt \bbt_{\rm{WN}}+\dt \bbt_{\rm{RN}}+\dt \bbt_{\rm{SGWB}}.
\n
The first term $\bbt_{\rm{TM}}$ accounts for the inaccuracies of timing model \citep{Chamberlin:2014ria}. It can be modeled as $\bbt_{\rm{TM}}=M \bep$, where $M$ represents the timing model design matrix, and $\bm{\epsilon}$ denotes a small offset vector indicating the disparity between true and estimated parameters. Essentially, $M \bep$ corresponds to the linear term in the Taylor expansion centered at the estimated timing parameter.
The second term $\bn{WN}$ representing the contribution from the time-independent white noise accounts for measurement uncertainties. It can be modeled by three parameters, with parameter EFAC  as a scale factor on the TOA uncertainties, parameter EQUAD as an added variance~\citep{NANOGrav:2015qfw} and parameter ECORR as a per-epoch variance for each backend/receiver system \citep{NANOGrav:2015aud}.  
The third term $\bn{RN}$ representing the stochastic contribution from time-correlated red noise accounts for the irregularities of the pulsar's motion~\citep{Shannon:2010bv}. It is modeled by a power-law spectrum with the amplitude $A_{\rm{RN}}$ and the spectral index $\gamma_{\rm{RN}}$. The last term $\bn{SGWB}$ is the contribution from an SGWB, which can be described by the cross-power spectral density \Eq{psd}. In practice, we adopt the ``Fourier-sum" method to calculate $\bn{RN}$ and $\bn{SGWB}$, choosing $N_{\text{mode}}$ discrete frequency modes as $f=1/T, 2/T, \dots, N_{\text{mode}}/T$ where $T$ is the observational timespan. Following \cite{NANOGrav:2023gor},  we choose $N_{\text{mode}}=30$ for the red noise of the individual pulsar and $N_{\text{mode}}=14$ for the common SGWB signal among all pulsars.

In the search for the SGWB from massive gravity, we adopt a Bayesian inference approach, following the methodology outlined in \cite{NANOGrav:2023gor}. The posterior distribution for the model parameter $\mathrm{\Theta}$ is given by
\m
P(\mathrm{\Theta}|\dt \bbt) \propto L(\dt \bbt|\mathrm{\Theta})\pi(\mathrm{\Theta}),
\n
where $L(\dt \bbt|\mathrm{\Theta})$ represents the likelihood evaluated by a multivariate Gaussian function \citep{Ellis:2013nrb} and $\pi(\mathrm{\Theta})$ denotes the prior distribution. The parameters and their prior distributions required for our analysis are detailed in Table \Table{prior}. In the analysis, we first infer the noise parameters for each individual pulsar without including the common signal $\dt \bbt_{\rm{SGWB}}$.
Then we combine all 67 pulsars as a collective unit, maintaining the white noise parameters at their maximum-likelihood values from the single-pulsar analysis, while allowing both the single-pulsar red noise and the common SGWB signal parameters to vary simultaneously.
We also assess the goodness-of-fit of two candidate hypotheses to the data by calculating the Bayes factor, defined as
\e
\BF \equiv \frac{\rm{Pr}(\dt \bbt|\mathcal{H}_2)}{\rm{Pr}(\dt \bbt|\mathcal{H}_1)},
\q
where $\rm{Pr}(\dt \bbt|\mathcal{H})$ measures the evidence that the data $\dt \bbt$ are produced under the hypothesis $\mathcal{H}$. Following the interpretation from \citep{BF}, when $\BF \le 3$, the evidence favoring $\mathcal{H}_2$ over $\mathcal{H}_1$ is deemed to be ``not worth more than a bare mention". However, the strength of evidence increases as $\BF$ grows, classified as ``substantial", ``strong" and ``decisive" when $3 \le \BF \le 10$, $10 \le \BF \le 100$, and $\BF \ge 100$, respectively. In practice, we estimate the Bayes factors using the \textit{product-space method} \citep{10.2307/2346151,10.2307/1391010,Hee:2015eba,Taylor:2020zpk}. 

All the above analyses are based on JPL solar system ephemeris (SSE) DE440 \citep{Park_2021}. To perform these Bayesian computations, we employ open-source software packages, namely \texttt{enterprise} \citep{enterprise} and \texttt{enterprise\_extension} \citep{enterprise_extensioins}, for likelihood and Bayes factor calculations. The Markov chain Monte Carlo (MCMC) sampling is facilitated by the \texttt{PTMCMCSampler} \citep{justin_ellis_2017_1037579} package.

\section{\label{sec:result}Results and discussion}

As we have demonstrated in \Sec{sec:orf}, an SGWB originating from massive gravity exhibits two fundamental distinctions when compared to one from massless gravity: the presence of a minimal frequency and a deviation in the Hellings-Downs ORF. Our graviton mass constraints are derived from an analysis of these two aspects.

Firstly, the NANOGrav collaboration analyzed their 15-year data set with a free spectrum, allowing for variations in the amplitudes of each individual Fourier frequency component. The presence of a non-vanishing amplitude at the lowest Fourier frequency illustrates the continued existence of a gravitational-wave signal at this frequency, implying that 
\m
f_{\text{min}}<1/T,
\n
which can be translated into the basic constraint on the graviton mass,
\m
m_g<8.2\times 10^{-24}\,\rm{eV}/c^2.
\n

\begin{figure}[htbp!]
	\centering
	\includegraphics[width=0.45\textwidth]{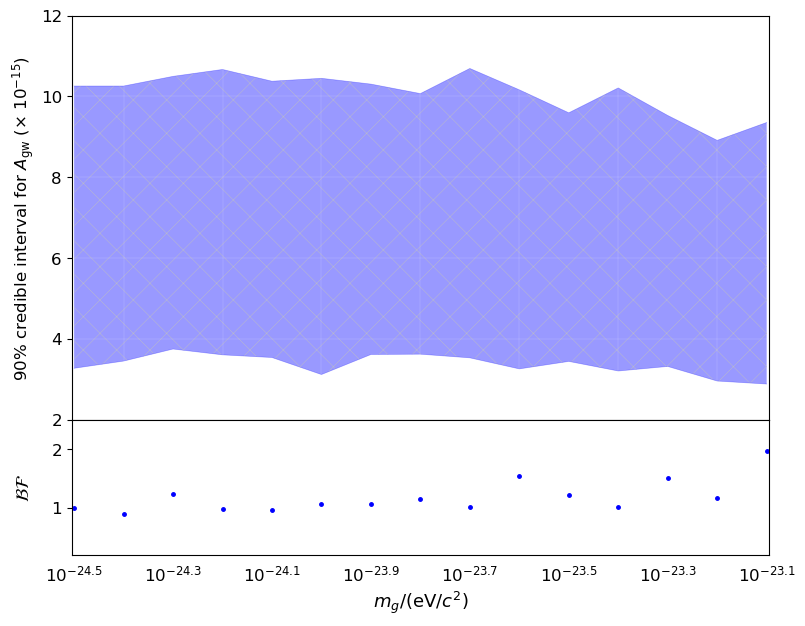}
	\caption{\label{A_BF} Top panel: the $90\%$ credible interval of the power spectrum amplitude $A_\mathrm{gw}$ as a function of the graviton mass $m_g$, from NANOGrav 15-year data set. Bottom panel: the corresponding Bayes factor, $\BF$, as a function of the graviton mass $m_g$.}
\end{figure}

Secondly, in the case of a lighter graviton mass, we calculate the Bayes factor between the hypothesis $H_2$ of a massive-gravity SGWB with correlations given in \Eq{orf2}, and the hypothesis $H_1$ of a massless-gravity SGWB with Hellings-Downs correlations. Our investigation spanned a graviton mass range of $m_g\in [3\times 10^{-25}, 8 \times 10^{-24}]\,\rm{eV}/c^2$. Here $3\times 10^{-25}\,\rm{eV}/c^2$ serves as an approximation for the massless scenario, as for $m_g\le 3\times 10^{-25}\,\rm{eV}/c^2$, $\eta>0.999$ across the all $14$ frequency bins. For each graviton mass listed in \Table{prior}, the corresponding Bayes factor is decipted in the bottom panel of \Fig{A_BF}. The results demonstrate that all Bayes factors are less than $3$ within the explored mass range. This suggests that it is challenging for current data to distinguish whether the SGWB arises from massive gravity or massless gravity based solely on spatial correlations. We also present the $90\%$ credible intervals for the power spectrum amplitude $A_{\rm{gw}}$ corresponding to each graviton mass in the top panel of \Fig{A_BF}. Within the mass range we explore, these $90\%$ credible intervals are remarkably consistent, with the majority of them falling within the range of $A_\mathrm{gw}\in [3,10]\times 10^{-15}$. These outcomes also closely align with the results of massless-gravity  SGWB reported in \citep{NANOGrav:2023gor}.


We note that \cite{Wang:2023div} also constrains the graviton mass through a distinct method. They transform the ORF into a function of the group speed $v/c=\sqrt{1-(\frac{m_g c^2}{\hbar \omega})^2}$ and evaluate its deviation from observed ORF with binned angular separation provided by \citep{NANOGrav:2023gor}. 
However, a challenge arises when translating the constraint on the group speed into a constraint on the graviton mass. This challenge stems from the difficulty in determining the appropriate frequency value (or the corresponding $\omega$) to use, as the observed ORF is obtained by assuming independence from frequency. 
In contrast, our search methodology takes a more straightforward approach by directly investigating a frequency-dependent ORF within the data set. 
Searching for the massive SGWB signal in the real data prohibits us to probe graviton mass larger than the lowest Fourier frequency $1/T$. However, the free spectrum have demonstrated the existence of gravitational-wave signal at this frequency. 
By combining this observation with the inherent existence of a minimum frequency in massive gravity, we obtain a natural upper bound for the graviton mass. This approach allows us to directly examine the frequency dependence and provides a meaningful constraint on the maximum graviton mass based on the available data.

\begin{acknowledgments}
We acknowledge the use of HPC Cluster of ITP-CAS. QGH is supported by the grants from NSFC (Grant No.~12250010, 11975019, 11991052, 12047503), Key Research Program of Frontier Sciences, CAS, Grant No.~ZDBS-LY-7009, the Key Research Program of the Chinese Academy of Sciences (Grant No.~XDPB15). 
ZCC is supported by the National Natural Science Foundation of China (Grant No.~12247176 and No.~12247112) and the China Postdoctoral Science Foundation Fellowship No.~2022M710429.
\end{acknowledgments}
\bibliographystyle{aasjournal}
\bibliography{massive_SGWB}

\end{document}